\documentclass[12pt]{iopart}

\begin{document}

\title[Minimum-error state discrimination]{On the conditions for discrimination between quantum states with minimum error}


\author{Stephen M. Barnett$^1$ and Sarah Croke$^2$}

\address{$^1$ Department of Physics, University of Strathclyde, Glasgow G4 0NG, UK}

\address{$^2$ Perimeter Institute for Theoretical Physics, Waterloo, Ontario, N2L 2Y5, Canada}
\ead{steve@phys.strath.ac.uk}
\begin{abstract}
We provide a simple proof for the necessity of conditions for discriminating with 
minimum error between a known set of quantum states.

\end{abstract}

\submitto{\JPA}

In quantum communications a transmitting party, Alice, selects from among a set of agreed quantum states
to prepare a quantum system for transmission to the receiving party, Bob.  Both the set of possible states,
$\{\hat\rho_i\}$, and the associated probabilities for selection, $\{p_i\}$ are known to Bob but not, of course,
the selected state.  His task is to determine as well as he can which state was prepared and he does this by
choosing a measurement to perform.  If the states are not mutually orthogonal then there is no measurement
that will reveal the selected state with certainty.  The strategy he chooses will depend on the use for which the 
information is intended and there exist many figures of merit for Bob's measurement \cite{Chefles00,Barnett08}.  Among these the simplest is the minimum probability of error or, equivalently, the maximum probability for correctly identifying the state.  Necessary and sufficient conditions for realising a minimum error measurement are known \cite{Holevo73,Yuen75,Helstrom76,Eldar03} but it has proven to be easier to prove sufficiency than necessity.  This letter presents an appealingly simple proof that the conditions are necessary.

A minimum error measurement will, in general, be a generalised measurement described not by projectors but
rather by a probability operator measure (POM) \cite{Helstrom76}, also referred to as a positive operator valued measure \cite{Peres93}.  The
probability that a generalised measurement gives the result $j$ is
\begin{equation}
\label{Eq1}
P(j) = {\rm Tr}\left(\hat\rho \hat\pi_j\right) ,
\end{equation}
where $\hat\pi_j$ is a probability operator.  These are defined, mathematically, by the requirements that
\begin{eqnarray}
\label{Eq2}
\hat\pi_i^\dagger &=& \hat\pi_i  \\
\label{Eq3}
\hat\pi_i &\ge & 0 \qquad {\rm or} \quad \langle \psi|\hat\pi_i|\psi\rangle  \ge 0
\quad \forall \: |\psi\rangle \\
\label{Eq4}
\sum_i\hat\pi_i &=& \hat{1} .
\end{eqnarray}
A minimum error measurement identifies the outcome $i$ with the prepared state $\hat\rho_i$ and the 
probability for correctly identifying the state is therefore
\begin{equation}
\label{Eq5}
P_{\rm corr} = \sum_ip_i{\rm Tr}\left(\hat\rho_i\hat\pi_i\right) ,
\end{equation}
and the error probability is, of course, $P_{\rm err} = 1-P_{\rm corr}$.

The conditions for minimum error are 
\begin{eqnarray}
\label{Eq6}
\hat\pi_j\left(p_j\hat\rho_j - p_k\hat\rho_k\right)\hat\pi_k &=& 0 \quad \forall j,k \\
\label{Eq7}
\sum_i p_i\hat\rho_i\hat\pi_i - p_j\hat\rho_j &\ge & 0 \quad \forall j . 
\end{eqnarray}
The latter condition further requires that the operator
\begin{equation}
\label{Eq8}
\hat\Gamma = \sum_i p_i\hat\rho_i\hat\pi_i
\end{equation}
must be Hermitian, for if it has an anti-Hermitian part then its expectation value can be
complex rather than the required real and positive value.  It is straightforward to show 
that the condition (\ref{Eq7}) is sufficient to minimise the error.  To see this let us 
consider another (primed) measurement associated with the POM $\{\hat\pi'_j\}$.  The
difference between the probabilities for correctly identifying the state with the minimum
error and primed measurements is
\begin{eqnarray}
\label{Eq9}
P_{\rm corr} - P'_{\rm corr} &=& \sum_ip_i{\rm Tr}\left(\hat\rho_i\hat\pi_i\right)
- \sum_jp_j{\rm Tr}\left(\hat\rho_j\hat\pi'_j\right) \nonumber \\
&=& \sum_j{\rm Tr}\left[\left(\hat\Gamma - p_j\hat\rho_j\right)\hat\pi'_j\right] \nonumber \\
&\ge & 0 ,
\end{eqnarray}
where we have used the completeness condition (\ref{Eq4}) for the primed probability operators
and the final inequality follows from the assumption that the original (unprimed) measurement
minimises the error probability.  The probability operators $\hat\pi'_j$ are positive by
virtue of the fact that they represent a measurement.  If the operators $\hat\Gamma - p_j\hat\rho_j$
are also positive then it follows immediately that 
${\rm Tr}\left[\left(\hat\Gamma - p_j\hat\rho_j\right)\hat\pi'_j\right] \ge 0$.  If we can
find a POM that satisfies the inequalities (\ref{Eq7}) then it will be a minimum error strategy.
This establishes the sufficiency of the condition (\ref{Eq7}).

In order to prove that (\ref{Eq7}) is also necessary we introduce the manifestly Hermitian operators
\begin{equation}
\label{Eq10}
\hat G_j = \frac{1}{2}\sum_ip_i\left(\hat\rho_i\hat\pi_i + \hat\pi_i\hat\rho_i\right) 
- p_j\hat\rho_j ,
\end{equation}
where the operators $\{\hat\pi_i\}$ comprise a minimum error measurement.  It is straightforward to
show that each of the operators $\hat G_j$ must be positive by considering the effects of a single
negative eigenvalue.  Let us suppose that for one state, $\hat\rho_1$, the operator $\hat G_1$ 
has a single negative eigenvalue, $-\lambda$:
\begin{equation}
\label{Eq11}
\hat G_1|\lambda\rangle = -\lambda|\lambda\rangle .
\end{equation}
If this single negative eigenvalue means that there exists a POM with a lower error probability then
it \emph{necessarily} follows that the positivity of $\hat G_1$ (and by extension of all of the 
operators $\hat G_j$) is a necessary condition for a minimum error POM.

Consider a measurement with probability operators $\hat\pi'_i$ related to the operators $\hat\pi_i$
by 
\begin{equation}
\label{Eq12}
\hat\pi'_i = \left(\hat{1} - \varepsilon|\lambda\rangle\langle\lambda|\right)
\hat\pi_i\left(\hat{1} - \varepsilon|\lambda\rangle\langle\lambda|\right)
+ \varepsilon(2-\varepsilon)|\lambda\rangle\langle\lambda|\delta_{i1} ,
\end{equation}
where the positive quantity $\varepsilon \ll 1$.  It is easily verified that the set
of these primed operators satisfies the conditions (\ref{Eq2})-(\ref{Eq4}) and so represents a 
valid measurement.  The probability that the primed measurement will correctly identify the
state is
\begin{eqnarray}
\label{Eq13}
P'_{\rm corr} &=& \sum_ip_i{\rm Tr}\left(\hat\rho_i\hat\pi'_i\right) \nonumber \\
&=& P_{\rm corr} - 
\varepsilon \sum_i\langle\lambda|\left(\hat\rho_i\hat\pi_i + \hat\pi_i\hat\rho_i\right)|\lambda\rangle
+ 2\varepsilon p_1\langle\lambda|\hat\rho_1|\lambda\rangle + O(\varepsilon^2) \nonumber \\
&=& P_{\rm corr} + 2\varepsilon\lambda + O(\varepsilon^2) ,
\end{eqnarray}
where we have used the eigenvalue property (\ref{Eq11}).  This is clearly greater than $P_{\rm corr}$
and so is at odds with the assumption that $P_{\rm corr}$ is the maximum probability for correctly
identifying the state.  It follows that the positivity of the operators $\hat G_j$ is a 
\emph{necessary} condition for maximising the probability of identifying the state or, equivalently,
for minimising the probability of error.  

We complete our proof of the necessity of the positivity condition (\ref{Eq7}) by showing that the
operator $\hat\Gamma$ must be Hermitian so that
\begin{equation}
\label{Eq14}
\hat G_j = \hat\Gamma - p_j\hat\rho_j .
\end{equation}
To see this we need only note that 
\begin{equation}
\label{Eq15}
\sum_j {\rm Tr}\left(\hat G_j \hat\pi_j\right) = 0 .
\end{equation}
Because both $\hat G_j$ and $\hat\pi_j$ are both positive operators it must then be the case that 
$\hat G_j\hat\pi_j = 0$.  Summing this over all $j$ then gives
\begin{equation}
\label{Eq16}
\frac{1}{2}\sum_i p_i \left(\hat\pi_i\hat\rho_i - \hat\rho_i\hat\pi_i \right) =
\frac{1}{2}\left(\hat\Gamma^\dagger - \hat\Gamma\right) = 0 ,
\end{equation}
so that the operator $\hat\Gamma$ is necessarily Hermitian.  This concludes the proof
of the necessity of the positivity condition (\ref{Eq7}) for any minimum error measurement.

We conclude by showing how the equality condition (\ref{Eq6}) follows from the inequality
condition (\ref{Eq7}).  The positivity of the operators $\hat\Gamma - p_j\hat\rho_j$ 
and $\hat\pi_j$ together with the trivial condition
\begin{equation}
\label{Eq17}
\sum_j{\rm Tr}\left[\left(\hat\Gamma - p_j\hat\rho_j\right)\hat\pi_j\right] = 0
\end{equation}
mean that
\begin{eqnarray}
\label{Eq18}
\left(\hat\Gamma - p_k\hat\rho_k\right)\hat\pi_k &=& 0 \\
\label{Eq19}
\hat\pi_j\left(\hat\Gamma - p_j\hat\rho_j\right) &=& 0 .
\end{eqnarray}
If we premultiply (\ref{Eq18}) by $\hat\pi_j$, postmultiply (\ref{Eq19}) by
$\hat\pi_k$ and take the difference then we recover the condition (\ref{Eq6}).
We conclude that each of the minimum error conditions (\ref{Eq6}) and (\ref{Eq7}) 
are both sufficient and necessary.  For any set of states and preparation probabilities
there will exist at least one minimum error measurement with probability operators 
satisfying these conditions.

\ack This work was supported by the Royal Society and the Wolfson foundation (SMB), and by Perimeter Institute for Theoretical Physics (SC). Research at Perimeter Institute is supported by the Government of Canada through Industry Canada and by the Province of Ontario through the Ministry of Research \& Innovation.


\section*{References}
\begin{thebibliography}{10}
\bibitem{Chefles00} Chefles A 2000 {\it Contemp. Phys.} \textbf{41} 401

\bibitem{Barnett08} Barnett S M and Croke S 2008 {\it Adv. Opt. Photonics} submitted

\bibitem{Holevo73} Holevo A S  1973 {\it J. Multivariate Analysis} \textbf{3} 337

\bibitem{Yuen75} Yuen H P, Kennedy R S and Lax M 1975 {\it IEEE Trans. Inf. Theory} \textbf{IT-21} 125

\bibitem{Helstrom76} Helstrom C W 1976 {\it Quantum detection and estimation theory} (New York: Academic Press)

\bibitem{Eldar03} Eldar Y C, Mergretski A and Verghese G C 2003 {\it IEEE Trans. Inf. Theory} \textbf{49} 1007

\bibitem{Peres93} Peres A 1993 {\it Quantum theory: concepts and methods} (Dordrecht: Kluwer)

\end{thebibliography}
\end{document}